\documentclass[%
preprint,
runinaddress,
 amsmath,amssymb,
 aps,
apl,
]{revtex4-1}

\usepackage{graphicx}
\usepackage{color}
\usepackage{dcolumn}
\usepackage{bm}
\usepackage{braket}

\begin{document}

\title{Densities of states in Fe-doped III-V semiconductors: a first-principles study}

\author{Shoya Sakamoto}
\affiliation{Department of Physics, The University of Tokyo, Bunkyo-ku, Tokyo 113-0033, Japan}

\author{Atsushi Fujimori}
\affiliation{Department of Physics, The University of Tokyo, Bunkyo-ku, Tokyo 113-0033, Japan}

\date{\today}
\begin{abstract}
The electronic structures of Fe-doped III-V semiconductors were studied by first-principles supercell calculation. 
It was found that their electronic structures are basically the same as those of Mn-doped ones except that the extra electron of Fe compared to Mn occupies either majority-spin $p$-$d$ hybridized antibonding states ($t_{a,\uparrow}$) or minority-spin $e$ states ($e_{\downarrow}$) and that the center of gravity of the $d$ partial density of states is higher for Fe than for Mn.
The present calculations suggest that ferromagnetism appears when the $e_{\downarrow}$ states start to be occupied.
The band splitting due to $s$-$d$ hybridization was found to be significantly smaller than the one due to $p$-$d$ hybridization. This indicates that the $s,p$-$d$ exchange interaction is not responsible for the high-temperature ferromagnetism of the Fe-doped ferromagnetic semiconductors even in $n$-type compounds.

\end{abstract}

\pacs{Valid PACS appear here}
\maketitle


\section{Introduction}
Ferromagnetic semiconductors (FMSs) have been studied extensively for decades since the discovery of Mn-doped III-V FMSs \cite{Munekata:1989aa,Ohno:1996aa,Jungwirth:2014aa,Dietl:2014aa,Tanaka:2014aa,Dietl:2015aa}. 
More recently, Fe-doped FMSs were synthesized \cite{Nam-Hai:2012ac,Tu:2014aa,Anh:2015aa,Tu:2018aa,Wakabayashi:2014ab} and have attracted much interest because they possess several advantages over the prototypical  (Ga,Mn)As or (In,Mn)As. First of all, the Curie temperatures ($T_{\rm C}$) are very high. Especially, the $T_{\rm C}$'s of (Ga,Fe)Sb and (In,Fe)Sb are 340 K \cite{Tu:2016aa} and 335 K \cite{Tu:2018aa} at highest, respectively, exceeding not only the highest $T_{\rm C} \sim 200$ K of (Ga,Mn)As \cite{Chen:2011aa} but also room temperature.
Second, various types of transport properties are realized: $p$-type semiconducting for (Ga,Fe)Sb \cite{Tu:2014aa} and Ge:Fe \cite{Ban:2014aa}, $n$-type semiconducting for (In,Fe)As:Be \cite{Nam-Hai:2012ac} and (In,Fe)Sb \cite{Tu:2018aa}, and insulating for (Al,Fe)Sb \cite{Anh:2015aa}. 
However, the basic electronic structure and the mechanism of the ferromagnetism as well as the origin of the diverse transport properties remain unclear and should be revealed for further development in this direction.

First-principles calculation based on density functional theory (DFT) is a powerful method for such a purpose, and has been widely used to study the electronic structures of FMSs \cite{Jungwirth:2006aa,Sato:2010aa}. Although it is difficult to exactly simulate randomly doped systems, a supercell approach provides the useful information about the basic electronic structures and the magnetic properties \cite{Sato:2000aa,Sandratskii:2002aa,Shick:2004aa,Wierzbowska:2004aa,Filippone:2011aa}. In the present study, we have constructed $3\times3\times3$ supercells and calculated the electronic structure or the density of states (DOS) of newly-discovered Fe-doped III-V FMSs as well as the spin and orbital magnetic moments.

\begin{figure}
\begin{center}
\includegraphics[width=8.0cm]{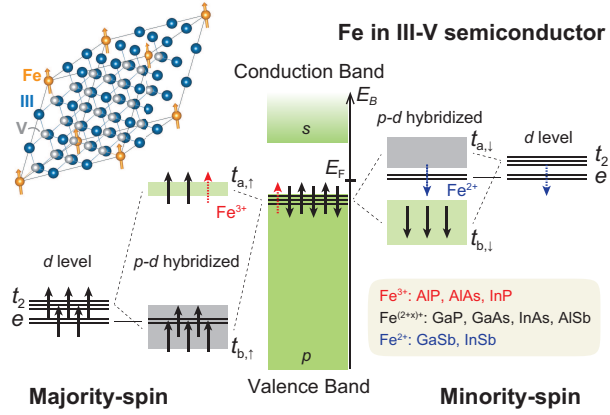}
\caption{Schematic energy diagram of Fe-doped III-V semiconductors. The host bands are shown in the middle by green boxes, and the majority-spin and minority-spin $3d$ ($t_{2}$ and $e$) levels of Fe atoms are shown on the left- and right-hand side, respectively. 
The $t_{2}$ orbitals and the host $p$ orbitals form bonding $t_{b}$ and anti-bonding $t_{a}$ orbitals through $p$-$d$ hybridization, which are shown by boxes between the host bands and the 3$d$ levels. Green and gray colors represent states with predominant $t_{2}$ and $p$ character, respectively.
Depending on the host semiconductor, Fe can take either Fe$^{3+}$ or Fe$^{2+}$ states, the electron occupancy of which is depicted by red and black arrows or blue and black arrows, respectively.
At the top left corner, the 3 $\times$ 3 $\times$ 3 supercell containing one Fe atom is also shown.}
\label{DFT_Supercells}
\end{center}
\end{figure}

\section{Methods}

\begin{table}[h!]
\begin{center}
\caption{Parameters used for the present supercell calculations: the lattice constants $a$ [\AA]; the muffin tin radii of Fe, group-III atoms, and group-V atoms, denoted by $R_{\rm MT}^{\rm Fe}$, $R_{\rm MT}^{\rm III}$, and $R_{\rm MT}^{\rm V}$ [Bohr]; the plane wave cutoff $K_{\rm max}$. Here, $R_{\rm MT}^{\rm min}$ denotes the radius of the smallest MT sphere.}
\begin{tabular}{p{5.5em}p{5em}p{3em}p{3em}p{3em}p{5em}}

\hline\hline
\\[-1em]
Unit cell \hfil & \hfil $a$ [\AA] \hfil & \hfil$R_{\rm MT}^{\rm Fe}$\hfil &  \hfil$R_{\rm MT}^{\rm III}$\hfil & \hfil$R_{\rm MT}^{\rm V}$\hfil & \hfil$R_{\rm MT}^{\rm min}K_{\rm max}$\hfil\\
\\[-1em]
\hline
${\rm Al_{26}FeP_{27}}$ & \hfil5.4672\hfil & \hfil2.5\hfil & \hfil2.33\hfil & \hfil1.95\hfil & \hfil5.5\hfil\\
${\rm Ga_{26}FeP_{27}}$ & \hfil5.4505\hfil & \hfil2.49\hfil & \hfil2.44\hfil & \hfil1.94\hfil & \hfil5.5\hfil\\
${\rm In_{26}FeP_{27}}$ & \hfil5.8697\hfil & \hfil2.5\hfil & \hfil2.5\hfil & \hfil2.09\hfil & \hfil5.5\hfil\\

${\rm Al_{26}FeAs_{27}}$ & \hfil5.6611\hfil & \hfil2.36\hfil & \hfil2.19\hfil & \hfil2.24\hfil & \hfil6.5\hfil\\
${\rm Ga_{26}FeAs_{27}}$ & \hfil5.6533\hfil & \hfil2.36\hfil & \hfil2.3\hfil & \hfil2.24\hfil & \hfil7.5\hfil\\
${\rm In_{26}FeAs_{27}}$ & \hfil6.0583\hfil & \hfil2.5\hfil & \hfil2.5\hfil & \hfil2.4\hfil & \hfil7.5\hfil\\

${\rm Al_{26}FeSb_{27}}$ & \hfil6.1355\hfil & \hfil2.49\hfil & \hfil2.3\hfil & \hfil2.49\hfil & \hfil6.5\hfil\\
${\rm Ga_{26}FeSb_{27}}$ & \hfil6.0959\hfil & \hfil2.47\hfil & \hfil2.41\hfil & \hfil2.47\hfil & \hfil7.5\hfil\\
${\rm In_{26}FeSb_{27}}$ & \hfil6.4794\hfil & \hfil2.5\hfil & \hfil2.5\hfil & \hfil2.5\hfil & \hfil8\hfil\\

\hline\hline

\end{tabular}
\label{DFT_param}
\end{center}
\vspace{0cm}
\end{table}

DFT calculations were done using the all-electron full-potential (linearized) augmented-plane-wave plus local-orbital method implemented in a WIEN2k package \cite{Blaha:2001aa}.
The spherical harmonic expansion was made up to $l = 10$ inside the muffin-tin spheres. The plane wave cutoff ($K_{\rm max}$) and the radii of MT spheres are summarized in Table \ref{DFT_param}.
For the exchange-correlation energy functional, the Perdew-Burke-Ernzerhof (PBE) generalized gradient approximation (GGA) was used \cite{Perdew:1996aa}.
In order to simulate the randomly doped system, supercells consisting of 3 $\times$ 3 $\times$ 3 primitive unit cell of the zinc blende structure are employed as shown at the top left corner of Fig. \ref{DFT_Supercells}, which contain 27 group-V atoms, 26 group-III atoms, and 1 Fe atom. 
This corresponds to $3.7\%$ Fe doping and would be a good starting point to study the basic electronic structure of isolated Fe atoms in III-V matrix without significant Fe-Fe interaction or hybridization. 
Brillouin-zone integration was performed on a 4 $\times$ 4 $\times$ 4 $k$-point mesh.
The experimental lattice constants of host semiconductors \cite{Vurgaftman:2001aa} were used for all the calculations for simplicity.
The self-consistent cycle was repeated until the calculated total energy converged to within 0.0001 Ry per supercell.

\section{Results}

\begin{figure*}
\begin{center}
\includegraphics[width=17.5cm]{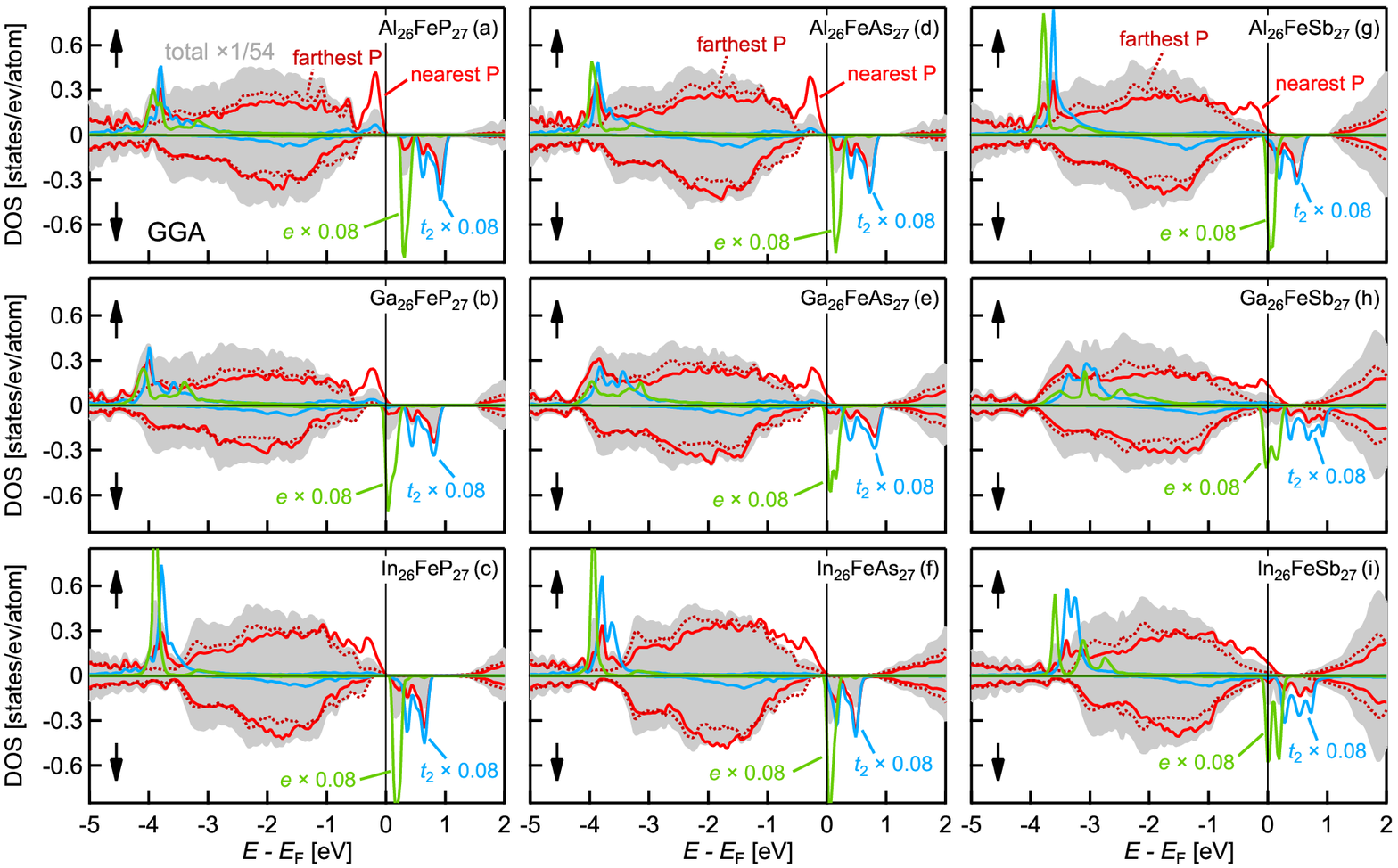}
\caption{Total and partial densities of states of Fe-doped III-V semiconductors, namely, (a) AlP, (b) GaP, (c) InP, (d) AlAs, (e) GaAs, (f) InAs, (g) AlSb, (h) GaSb, and (i) InSb. The majority-spin and minority-spin DOSs are plotted on the positive and negative sides of each graph.
Here, gray areas indicate the total DOSs, and red solid and dashed curves the PDOSs of the group-V atoms nearest to and farthest from the Fe atom, respectively. The PDOSs of the Fe $t_{2}$ and $e$ orbitals are also shown by blue and green curves. For ease of comparison, the total DOSs were divided by 54, and the Fe $t_{2}$ and $e$ PDOSs were multiplied by 0.08.}
\label{DOS}
\end{center}
\end{figure*}

Figure \ref{DOS} shows the spin-resolved total DOSs and partial DOSs (PDOSs) of Fe-doped III-V semiconductors. 
The total DOSs are indicated by gray areas, and the PDOSs of the group-V atoms located nearest to and farthest from the Fe atom are shown by red solid and dashed curves, respectively.
The PDOSs of Fe $t_{2}$ ($d_{xy}$,$d_{yz}$,$d_{zx}$) and $e$ ($d_{x^2-y^2}$,$d_{z^2}$) orbitals are also shown by blue and green curves. 
For the sake of easy comparison, the total DOS is divided by 54 (the total number of atoms in the supercell), and the PDOSs of $t_{2}$ and $e$ orbitals are multiplied by 0.08. Note that the (P)DOSs of majority-spin and minority-spin states are shown at the positive and negative sides of each panel.

Reflecting the calculated DOSs, the schematic energy diagram is illustrated in Fig. \ref{DFT_Supercells}. 
The valence and conduction bands, mainly consisting of the $p$ orbital of the group V element and the $s$ orbital of group III element, respectively, are shown in the middle of the figure by green boxes, while the majority-spin and minority-spin 3$d$ levels are shown on the left- and right-hand side, respectively. 
The 3$d$ levels split into doubly-degenerate lower $e$ state and triply-degenerate higher $t_{2}$ state because of the tetrahedral crystal field at the substitutional sites of group-III element in the zinc-blende crystal.
Furthermore, the $t_{2}$ orbitals and the ligand $p$ orbitals strongly hybridize with each other and form bonding and antibonding orbitals.
The majority-spin bonding states ($t_{b,\uparrow}$) of predominant $t_{2}$ character are located deep about 3-4 eV below the Fermi energy ($E_{\rm F}$), while the antibonding states ($t_{a,\uparrow}$) of predominant $p$ character are near the valence-band maximum. On the other hand, the minority-spin bonding states ($t_{b,\downarrow}$) mainly consisting of the ligand $p$ orbitals form part of the valence band, and antibonding states ($t_{a,\downarrow}$) of $t_{2}$ character remain unoccupied. Note that the $e$ orbitals do not strongly hybridize with the ligand orbitals.

In the case of the prototypical Mn-doped III-V systems, the $t_{a,\uparrow}$ level crosses the $E_{\rm F}$ accommodating a hole, because the substitution of Mn with the $3d^{5}(4sp)^{2}$ configuration for group III element with $(4sp)^{3}$ makes one electron missing from the valence band.
On the other hand, in the case of the Fe-doped III-V system, the $t_{a,\uparrow}$ states would be fully occupied since Fe has one more electron than Mn and can take the $3d^{5}(4sp)^{3}$ configuration.
In the present calculation, this is the case for (Al,Fe)P, (Al,Fe)As, and (In,Fe)P. 
To be precise, the Fe and the ligand orbitals take the $e_{\uparrow}^{2}(t_{b,\uparrow}t_{b,\downarrow})^{3}t_{a,\uparrow}^{3}$ configuration in those cases, while Mn in III-V semiconductors takes the $e_{\uparrow}^{2}(t_{b,\uparrow}t_{b,\downarrow})^{3}t_{a,\uparrow}^{2}$ configuration. 
(Ga,Fe)Sb and (In,Fe)Sb, however, showed distinct behavior from this.
That is, Fe and the ligand take the $e_{\uparrow}^{2}(t_{b,\uparrow}t_{b,\downarrow})^{3}t_{a,\uparrow}^{2}e_{\downarrow}^{1}$ configuration, where an electron occupies the $e_{\downarrow}$ state instead of the $t_{a,\uparrow}$. 
This is probably because the valence band or the $p$ level of GaSb and InSb lies rather high in energy \cite{Vurgaftman:2001aa}, and it becomes more stable for an electron to occupy the $e_{\downarrow}$ than to occupy the $t_{a,\uparrow}$ states.
The former electronic configuration $e_{\uparrow}^{2}(t_{b,\uparrow}t_{b,\downarrow})^{3}t_{a,\uparrow}^{3}$ is denoted as Fe$^{3+}$ hereafter, and the latter configuration $e_{\uparrow}^{2}(t_{b,\uparrow}t_{b,\downarrow})^{3}t_{a,\uparrow}^{2}e_{\downarrow}^{1}$ as Fe$^{2+}$.
The other systems such as (Ga,Fe)As and (Al,Fe)Sb lie between Fe$^{3+}$ and Fe$^{2+}$, i.e., a small amount of electrons occupies the $e_{\downarrow}$ states and a small amount of holes are introduced to the $t_{a,\uparrow}$ states accordingly.
This is represented by red and blue dashed arrows in Fig. \ref{DFT_Supercells}, where the electron of highest energy occupies either $t_{a,\uparrow}$ or $e_{\downarrow}$ levels.

\begin{figure}
\begin{center}
\includegraphics[width=8.2cm]{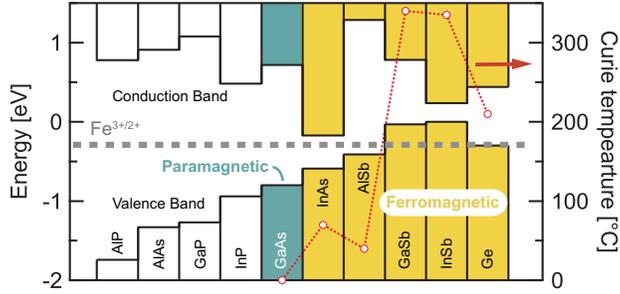}
\caption{Band offsets of III-V semiconductors \cite{Vurgaftman:2001aa} and Ge \cite{Franciosi:1996aa} with respect to the valence-band maximum of InSb. 
The boxes at the bottom represent the valence bands, while the ones at the top the conduction bands. The dashed line approximately represents the energy level above which Fe takes the Fe$^{2+}$ valence state rather than Fe$^{3+}$. The position of the line has been determined so that it corresponds to the Fe$^{3+/2+}$ charge transfer level in GaAs \cite{Malguth:2008aa}.}
\label{BandAlignment}
\end{center}
\end{figure}

This situation is illustrated in Fig. \ref{BandAlignment}, where the band offsets of III-V semiconductors \cite{Vurgaftman:2001aa} and Ge \cite{Franciosi:1996aa} are shown. 
The dashed line approximately represents the energy level above which the Fe$^{2+}$ configuration is stabilized. For example, in GaAs, the position of the dashed line shows the Fe$^{3+/2+}$ charge transfer level \cite{Malguth:2008aa}. 
The highest $T_{C}$ ever achieved for each material is also plotted. Considering that (Ga,Fe)As is paramagnetic \cite{Haneda:2000aa, Moriya:2001aa}, while the other Fe-doped semiconductors whose valence band maximum is located higher in energy than that of GaAs are ferromagnetic, ferromagnetism might appear when the Fe $e_{\downarrow}$ level starts being occupied.

The strength of the $p$-$d$ exchange or the spin splitting of the valence band $\Delta E_{v}$ caused by $p$-$d$ hybridization can be approximated as $t_{pd}^2/(E_{p}^{\uparrow}-E_{d}^{\uparrow})$, where $t_{pd}$ denotes the transfer integral between the ligand $p$ and the Fe $t_{2}$ orbitals, and $E_{p}^{\uparrow}$ and $E_{d}^{\uparrow}$ the energy level of the majority-spin ligand $p$ and the Fe $3d$ orbitals without the hybridization. 
For example, as the host semiconductor changes from AlP $\rightarrow$ GaP $\rightarrow$ InP, the $p$-$d$ hybridization becomes weaker because the increase in the lattice constant leads to the decrease in $t_{pd}$ and the higher energy position of the valence band ($E_{p}$) results in the increase in $E_{p}^{\uparrow}-E_{d}^{\uparrow}$.
This trend also holds for AlAs $\rightarrow$ GaAs $\rightarrow$ InAs series and basically for AlSb $\rightarrow$ GaSb $\rightarrow$ InSb series, too.
Table \ref{DFT_moments} summarizes the spin splitting of the valence band ($\Delta E_{v}$) and that of the conduction band ($\Delta E_{c}$) obtained as  the leading-edge difference between the majority- and minority-spin valence-band and conduction-band DOSs.
In fact, $\Delta E_{v}$ decreased as (Al,Fe)X $\rightarrow$ (Ga,Fe)X $\rightarrow$ (In,Fe)X, where X denotes P, As, or Sb.

The spin splitting of the conduction band $\Delta E_{c}$ is significantly smaller than that of the valence band since $s$-$d$ hybridization is very weak compared to the $p$-$d$ hybridization. A recent tunneling spectroscopy study \cite{Le-Duc-Anh:2016aa} on ferromagnetic (In,Fe)As revealed that the splitting of the conduction band was 31.7 and 50 meV for 6\% and 8\% Fe doping, respectively. These values agree with the calculated one of 20 meV for 3.7\% Fe doping, because the $N_{0}\alpha = \Delta E_{c}/x\braket{S}$ is calculated to be 0.22 eV assuming $\braket{S} = 5/2$ and is almost the same as the experimental 0.21-0.25 eV.

Table \ref{DFT_moments} also shows the calculated spin and orbital magnetic moments of Fe atom, and the total magnetic moment of the 3$\times$3$\times$3 supercell.
Because of the strong $p$-$d$ hybridization, a sizable amount of 3$d$ electrons also occupies $t_{b,\downarrow}$ states.
This results in the smaller magnetic moment of Fe atoms than the ionic value of 5 $\mu_{\rm B}$. 
Nevertheless, in the system with the empty $e_{\downarrow}$ level such as (Al,Fe)P or (In,Fe)P, the total magnetic moment of the supercell is exactly 5 $\mu_{\rm B}$. 
Similar behavior was also reported for (Ga,Mn)As \cite{Shick:2004aa}, where the DFT calculation based on the local spin-density approximation (LSDA) yielded the Mn moment of 3.47 $\mu_{\rm B}$, while the total magnetic moment in the supercell is exactly 4 $\mu_{\rm B}$.
In the case of (Ga,Fe)Sb and (In,Fe)Sb, where an extra electron occupies the $e_{\downarrow}$ states instead of the $t_{a, \uparrow}$, the total magnetic moment in the supercell is reduced by $\sim$1 $\mu_{B}$.
The intermediate systems such as (Ga,Fe)As or (Al,Fe)Sb showed the total magnetic moment slightly smaller than 5 $\mu_{B}$.
Note that the number of 3$d$ electrons is 6 for all the systems studied in the present work.

The orbital magnetic moment relative to the spin magnetic moment ($m_{l}/m_{s}$) increases from 0.015 to 0.061 as a host semiconductor becomes heavier and spin-orbit interaction becomes stronger.
This is consistent with the previous x-ray magnetic circular dichroism (XMCD) experiments, where $m_{l}$/$m_{s}$ is 0.1 for (In,Fe)As \cite{Sakamoto:2016aa} and 0.13 for (Ga,Fe)Sb \cite{Sakamoto:2019ac}.

\begin{table}
\begin{center}
\caption{Spin and orbital magnetic moments of Fe atom, spin magnetic moment in a unit cell, the spin splitting of the valence band and the conduction band.}
\begin{tabular}{p{5.0em}p{4em}p{4em}p{4em}p{3.5em}p{3.5em}}
\hline\hline
\\[-1em]
material \hfil & \hfil $m_{s}$ \hfil & \hfil$m_{l}$/$m_{s}$\hfil &  \hfil$M_{s}$\hfil & \hfil$\Delta E_{v}$\hfil & \hfil$\Delta E_{c}$\hfil\\
& \hfil [$\mu_{\rm B}$/Fe] \hfil &  &  \hfil [$\mu_{\rm B}$/cell]\hfil & \hfil[eV]\hfil & \hfil[eV]\hfil\\
\\[-1em]
\hline
(Al,Fe)P & \hfil3.33\hfil & \hfil0.015\hfil & \hfil5.00\hfil & \hfil0.41\hfil & \hfil0.08\hfil\\
(Ga,Fe)P & \hfil3.31\hfil & \hfil0.016\hfil & \hfil4.91\hfil & \hfil0.32\hfil & \hfil0.04\hfil\\
(In,Fe)P & \hfil3.51\hfil & \hfil0.017\hfil & \hfil5.00\hfil & \hfil0.21\hfil & \hfil0.02\hfil\\

(Al,Fe)As & \hfil3.33\hfil & \hfil0.023\hfil & \hfil4.97\hfil & \hfil0.34\hfil & \hfil0.06\hfil\\
(Ga,Fe)As & \hfil3.37\hfil & \hfil0.024\hfil & \hfil4.76\hfil & \hfil0.23\hfil & \hfil0.03\hfil\\
(In,Fe)As & \hfil3.48\hfil & \hfil0.027\hfil & \hfil4.96\hfil & \hfil0.22\hfil & \hfil0.02\hfil\\

(Al,Fe)Sb & \hfil3.29\hfil & \hfil0.043\hfil & \hfil4.89\hfil & \hfil0.28\hfil & \hfil0.04\hfil\\
(Ga,Fe)Sb & \hfil3.06\hfil & \hfil0.049\hfil & \hfil3.69\hfil & \hfil0.34\hfil & \hfil0.04\hfil\\
(In,Fe)Sb & \hfil3.30\hfil & \hfil0.061\hfil & \hfil4.16\hfil & \hfil0.25\hfil & \hfil0.03\hfil\\

\hline\hline

\end{tabular}
\label{DFT_moments}
\end{center}
\vspace{0cm}
\end{table}

\section{Discussion}

There are three kinds of exchange interactions often discussed in the field of dilute ferromagnetic semiconductors, namely, $s$,$p$-$d$ exchange interaction, double-exchange interaction, and superexchange interaction. 
$s$,$p$-$d$ exchange interaction, which has been applied to the Mn-doped FMSs, does not seem very important for the Fe-doped FMSs for the following reasons. First, the Curie temperatures of $n$-type (In,Fe)As \cite{Nam-Hai:2012ac} and (In,Fe)Sb \cite{Tu:2018aa} are similar to, or even higher than, that of $p$-type (Ga,Fe)Sb \cite{Tu:2015aa} despite the fact that $s$-$d$ exchange interaction is an order of magnitude weaker than $p$-$d$ exchange interaction as seen from Table \ref{DFT_moments}, which indicated that $\Delta E_{v} \gg \Delta E_{c}$. Second, ferromagnetism with $T_{\rm C}$ = 40 K was also reported for insulating (Al,Fe)Sb, whose carrier concentration was 3 $\times$ 10$^{17}$ cm$^{-3}$, three to four orders of magnitude smaller than that of (Ga,Mn)As \cite{Anh:2015aa}. Third, it was recently shown that the ferromagnetism of (In,Fe)Sb is not significantly influenced by carrier concentration and even by carrier type \cite{Kudrin:2019ac}.

Superexchange interaction is very short-ranged and antiferromagnetic for the nearest-neighbor Fe pairs and is often treated as obstacles for ferromagnetism in FMSs. Shinya $et$ $al$. \cite{Shinya:2018aa} pointed out that the second-nearest-neighbor superexchange interaction is actually ferromagnetic for (Ga,Fe)Sb and (In,Fe)Sb, although the magnitude of the interaction was too small to account for their high $T_{\rm C}$s.

If the Fe$^{2+}$ state is realized, there can also be double-exchange interaction between partly-filled $e_{\downarrow}$ orbitals. Such a scenario was proposed in the theoretical calculation on (In,Fe)As:Be done by Vu $et$ $al$. \cite{Vu:2014aa}, where they claimed that extra electrons introduced by Be doping occupy the $e_{\downarrow}$ orbitals and can induce a ferromagnetic order. The fact that the $T_{\rm C}$ tends to be higher in Sb based material, where the calculation yielded the Fe$^{2+}$ state rather than Fe$^{3+}$, may suggest that double-exchange interaction is more likely to be responsible for the ferromagnetic order. 

It is worth mentioning that there might be a long-range exchange interaction between the $e_{\downarrow}$ orbitals and the host bands both in $p$- and $n$-type semiconductors especially if they have narrow band gaps \cite{Gu:2016aa}. Such an interaction might be resonantly enhanced when $e_{\downarrow}$ orbitals and host bands are close in energy as suggested by Hai {\it et al.} \cite{Nam-Hai:2012ab} and may also play a role in stabilizing long-range ferromagnetic order.

Considering the low concentrations of Fe atoms doped into the host semiconductors, it appears necessary to think about the inhomogeneous distribution of Fe atoms on the nanoscale, or the spinodal nanodecomposition \cite{Dietl:2015aa}, because both superexchange and double-exchange interactions are short-ranged.
In fact, previous XMCD measurements on (In,Fe)As:Be \cite{Sakamoto:2016aa} and (Al,Fe)Sb \cite{Sakamoto:2019ab} thin films grown by the molecular beam epitaxy methods revealed that there exist nanoscale ferromagnetic domains even at room temperature much above the macroscopic $T_{\rm C}$, the origin of which was attributed to the nanoscale Fe concentration fluctuation. Furthermore, the nanoscale Fe-rich lamellae-like structures were recently observed in (In,Fe)As thin films prepared by the pulsed laser melting method \cite{Yuan:2018aa}. From theoretical points of view, it has been shown that it is more energetically stable for Fe atoms to be distributed close to each other in (In,Fe)As:Be \cite{Yuan:2018aa} (especially in the presence of interstitial Be atoms \cite{Vu:2014aa}), (Ga,Fe)Sb, and (In,Fe)Sb \cite{Shinya:2018aa, Fukushima:2019aa}.
Note that, in every study mentioned above, Fe atoms do not precipitate but maintain the zinc-blende structure.
When such Fe-rich regions are formed and become locally highly metallic, it may be even possible that Stoner ferromagnetism emerges in those regions.

The transport property would be another puzzle if one assume the Fe$^{2+}$ configuration. This should introduce as many hole carriers as doped Fe atoms, but in reality, the carrier concentration is orders of magnitude smaller than the doped Fe atoms. This discrepancy would be resolved if one assumes that the carriers are trapped inside the Fe-rich regions and macroscopic carrier transport occurs via hopping between those Fe-rich regions. Such a model was originally introduced by Kaminski and Das Sarma for (Ga,Mn)As \cite{Kaminski:2003aa} and subsequently applied to Ge:Mn \cite{Pinto:2005aa}, Ge:Fe \cite{Ban:2018aa}, and (Zn,Cr)Te \cite{Sreenivasan:2007aa} to describe their insulating/semiconducting natures and measured low carrier concentrations ($\sim$10$^{18}$ cm$^{-3}$ for Ge:Mn \cite{Pinto:2005aa} and Ge:Fe \cite{Ban:2014aa},  $\sim$10$^{15}$ cm$^{-3}$ for (Zn,Cr)Te \cite{Saito:2002aa}).

We have not calculated how interstitial Fe atoms alter the electronic structure, but there would be two major effects. First, the electron carrier concentration would increase because interstitial Fe atoms would act as double/triple donors. Second, there might be a reduction in the net magnetization because interstitial Fe atoms would probably be antiferromagnetically coupled to adjacent Fe atoms. 
Although such effects should be rigorously assessed both theoretically and experimentally in future studies, they may be ignored in the first approximation considering orders of magnitude lower carrier concentration than the number of doped Fe atoms.


\section{Conclusion}

We have calculated the basic electronic structures of Fe-doped III-V semiconductors by first-principles supercell calculation. They were found to be similar to those of Mn-doped counterparts except that one more electron of Fe than Mn occupies either majority-spin antibonding states ($t_{a,\uparrow}$) or minority-spin $e$ states ($e_{\downarrow}$) and that the center of gravity of the $d$ partial density of states is higher for Fe than for Mn. The $e_{\downarrow}$ is preferentially occupied in the cases of (Ga,Fe)Sb and (In,Fe)Sb, where the $p$ level or the valence band is located high in energy and, therefore, it is more stable for the electron to occupy $e_{\downarrow}$ states than $t_{a,\uparrow}$ states. 
As the group-III element changes from Al $\rightarrow$ Ga $\rightarrow$ In, $p$-$d$ exchange interaction gets weaker because the increase of Fe-V bond length leads to the decrease of the transfer integral $t_{pd}$ between the Fe $t_{2}$ orbitals and ligand $p$ orbitals.
As the group-V element changes from P $\rightarrow$ As $\rightarrow$ Sb, $p$-$d$ exchange interaction also becomes weaker because more covalent nature and higher energy position of $p$ level makes $E_{p}^{\uparrow}-E_{d}^{\uparrow}$ larger.
The present calculations implied that the ferromagnetism originates from the nanoscale fluctuation of Fe atom distribution, otherwise the small amount of carriers is not likely to stabilize the long-range ferromagnetic order especially in the case of $n$-type (In,Fe)As and (In,Fe)Sb and insulating (Al,Fe)Sb.


\section*{acknowledgments}
This work was supported by Grants-in-Aids for Scientific Research from the JSPS (No. 15H02109).
A.F. is an adjunct member of the Center for Spintronics Research Network (CSRN), the University of Tokyo, under Spintronics Research Network of Japan (Spin-RNJ).
S.S. acknowledges financial support from Advanced Leading Graduate Course for Photon Science (ALPS) and the JSPS Research Fellowship for Young Scientists. 

\bibliography{BibTex_Fe_IIIV_DOS.bib}

\end{document}